\documentclass[prl,twocolumn,showpacs,amsmath,amssymb,superscriptaddress]{revtex4}
\usepackage{graphicx}
\usepackage{dcolumn}
\usepackage{bm}
\usepackage{epsfig}

\begin{document}


\title{Muon-fluorine entangled states in molecular magnets}

\author{T. Lancaster}
\email{t.lancaster1@physics.ox.ac.uk} 
\author{S.J. Blundell} 
\author{P.J. Baker}
\author{M.L. Brooks}
\author{W. Hayes}
\affiliation{Clarendon Laboratory, Oxford University Department of Physics, Parks
Road, Oxford, OX1 3PU, UK
}

\author{F.L. Pratt}
\affiliation{ISIS Facility, Rutherford Appleton Laboratory, Chilton, 
Oxfordshire OX11 0QX, UK}

\author{J.L. Manson}
\author{M.M. Conner}
\affiliation{Department of Chemistry and Biochemistry,
Eastern Washington University,
Cheney, WA 99004, USA
}
\author{J.A. Schlueter}
\affiliation{Material Science Division, Argonne National Laboratory, 
Argonne, IL 60439, USA}

\begin{abstract}
The information accessible from 
a muon-spin relaxation experiment is often limited
since we lack knowledge of the precise muon stopping site.
We demonstrate here the possibility of localizing a spin polarized
muon in a known stopping state in a molecular material containing
fluorine. The muon-spin precession that results from the
entangled nature of the muon-spin and surrounding nuclear
spins is sensitive
to the nature of the stopping site and we use this property
to identify three classes of site. We are also able to describe the
extent to which the muon distorts its surroundings.  
\end{abstract}
\pacs{76.75.+i, 75.50.Xx, 61.18.Fs}

\maketitle

\begin{figure}
\begin{center}
\epsfig{file=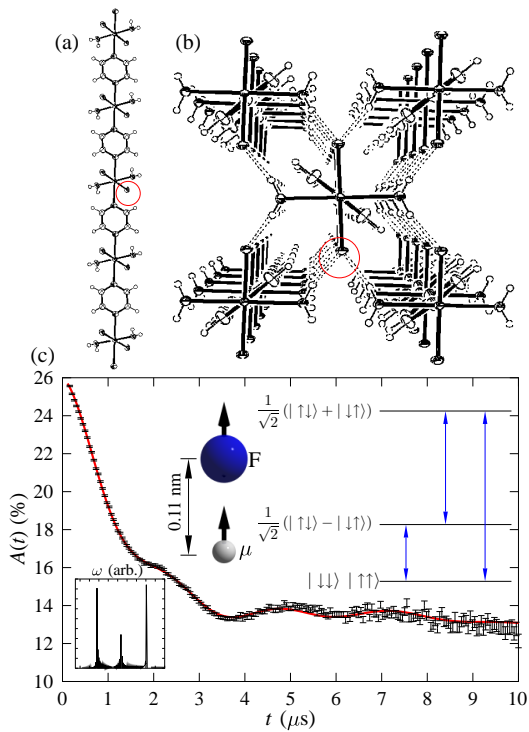,width=7.cm}
\caption{(Color online) (a) Linear chain structure of
  CuF$_{2}$(H$_{2}$O)$_{2}$(pyz). The circled regions shows 
sample fluorine atoms near which the muon is expected to localize.
(b) Crystal packing viewed parallel to the chain axis. Dotted lines
show F H--O hydrogen bonds.
(c) Example ZF $\mu^{+}$SR spectrum measured at $T=3.4$~K with a fit
to the F--$\mu^{+}$ dipole interaction described in the text. {\it
  Inset:} F--$\mu^{+}$ configuration, (top right)
the resulting energy levels with
allowed transitions and (bottom left) the calculated frequency spectrum.
\label{cuf2}}
\end{center}
\end{figure}

Muon-spin relaxation ($\mu^{+}$SR) continues to provide insights
into the nature of magnetic materials, superconductors, semiconductors, soft
matter and chemical reactions \cite{steve}. 
The technique involves stopping spin polarized muons in a target
sample where the muons probe the
distribution of local magnetic fields across the muon stopping sites. 
Despite its successes, two concerns are sometimes
raised. The first is
that the exact muon stopping site in a material is, in general, 
not precisely
known, introducing some uncertainty in the determination of the 
local spin structure. 
This is especially applicable in anisotropic or 
chemically complex materials where a number of different candidate
stopping sites exist.
A second objection is that it is not possible to quantify the degree to which
the presence of the charged muon distorts its surroundings and hence
to estimate the modification of superexchange pathways
that could, in principle, affect the local spin structure. 
Here we show that in addition to the
magnetic behavior usually probed by $\mu^{+}$SR, the entangled states
of $\mu^{+}$ spin and $^{19}$F nuclear spins allow the accurate determination 
of the muon site in a class of novel fluorinated molecular magnets,
along with an estimate of the distortion introduced by the presence of
the probe particle. 

Localized muons in solids often interact with their
environment via dipole-dipole coupling. In many cases the large number of
spin centres surrounding the muon allow the use of the
local magnetic field (LMF) approximation, where the muon
spin $S$ interacts with the net local magnetic field at the muon
site $\langle B \rangle$ via the Hamiltonian 
$\mathcal{H} = \gamma_{\mu}S \cdot \langle B \rangle$, 
where $\gamma_{\mu}$ is the muon gyromagnetic ratio.
For the commonly encountered case of the muon response to an ensemble
of randomized static local fields, the LMF model
gives the Kubo-Toyabe function \cite{hayano}, which is well
approximated by a Gaussian function $\exp(-\sigma^{2}t^{2})$ at early
times. Occasionally, however,  the muon is found to interact 
strongly with a small
number of spin centres through dipole coupling \cite{lord,celio}, a
process which is described by the Hamiltonian
\begin{equation}
\label{hamiltonian}
\mathcal{H} = \sum_{i > j} \frac{\mu_{0} \gamma_{i} \gamma_{j}}{4 \pi
  r^{3}} 
\left[ \mathbf{S}_{i} \cdot \mathbf{S}_{j} - 
3 (\mathbf{S}_{i} \cdot \hat{\mathbf{r}})(\mathbf{S}_{j} \cdot \hat{\mathbf{r}}  ) \right],
\end{equation}
where $\mathbf{r}$ is the vector linking spins $S_{i}$ and $S_{j}$, which have
gyromagnetic ratios $\gamma_{i,j}$. 
This strong interaction is found most often in materials containing
fluorine, for two reasons: fluorine is the most electronegative
element, causing muons to localise in its vicinity and 
fluorine occurs with a single isotope ($^{19}$F) with an $I=1/2$
nuclear spin, giving rise to a strong $\mu^{+}$SR signal
as a result of the interaction (see below). 
For example, in most insulating metal fluorides \cite{brewer} 
the muon sits midway between
 two fluorine ions
forming a strong linear ``hydrogen bond'' with an F--F
separation of $d=0.238$~nm, which is approximately twice the 
fluorine ionic radius. This stopping state (the so-called
F--$\mu^{+}$--F) is similar
to the $V_{\mathrm{k}}$ centre observed in alkali halides, which is
often treated as a molecule-in-a-crystal defect \cite{hayes}, where
the host weakly perturbs the molecular ion. 

The fact that the spins are not initially in an eigenstate of
the Hamiltonian in Eq.~(\ref{hamiltonian}) causes a time dependence of
the system's total
wavefunction, visualised classically as spontaneous precession of all 
spin species.
The observed property in $\mu^{+}$SR is the polarization
$D_{z}(t)$ of the
muon ensemble along a quantization axis $z$, which is given by \cite{roduner}
\begin{equation}
\label{polarization}
D_{z}(t) = \frac{1}{N} \left[ \sum_{m,n} | \langle m |\sigma_{z}| n \rangle | 
^{2} \cos(\omega_{mn}t)\right],
\end{equation}
where $N$ is the number of spins, 
$|m \rangle$ and $|n \rangle$ are eigenstates of the total Hamiltonian
$ \mathcal{H}$ and $\sigma_{z}$ is the Pauli spin matrix
corresponding to the quantization direction. 

These considerations suggest that, in compounds containing fluorine
ions, muons might be localized in known positions near fluorines.
The muon and fluorine spins then become entangled causing the
spins to evolve via the Hamiltonian in Eq.(\ref{hamiltonian}) and the
sensitivity of the interaction to the relative positions of the spin
centres allows the spin configuration and stopping site to be determined. 
This is indeed the case in a novel class of fluorine-containing
molecular magnets. The stopping states, however, are more complex than
the linear F--$\mu^{+}$-F molecular ion previously observed in many
inorganic insulating fluorides \cite{brewer}.

Molecular magnets are self assembled materials which 
are formed through bridging 
paramagnetic cation centres
(such as Cu$^{2+}$ in the compounds studied here (see Table
\ref{table})) 
with organic ligands.
These materials often possess low-dimensional (i.e.\ 2D, 1D)
structural motifs
and correspondingly display low-dimensional magnetic properties
\cite{steveandfrancis}.
Our recent experiments on such systems have shown that muons
are uniquely sensitive to the presence of long-range magnetic order
\cite{lancaster}. 
Below the antiferromagnetic
transition at $T_{\mathrm{N}}$ 
we observe oscillations in the time dependence of the muon
polarization (the ``asymmetry'' $A(t)$ \cite{steve}) which are
characteristic of a quasi-static local magnetic field at the 
muon stopping site. In the LMF model, this local field causes a 
coherent precession of the
spins of those muons for which a component of their spin polarization
lies perpendicular to this local field (expected to be 2/3 of the
total spin polarization for a powder sample). 
The frequency of the oscillations is given by
$\nu_{i} = \gamma_{\mu} |B_{i}|/2 \pi$, where $\gamma_{\mu}$ is the muon
gyromagnetic ratio ($=2 \pi \times 135.5$~MHz T$^{-1}$) and $B_{i}$
is the average magnitude of the local magnetic field at the $i$th muon
site. 
Above $T_{\mathrm{N}}$ the character of the measured spectra changes
considerably and we observe lower 
frequency oscillations
characteristic of the dipole interaction of the muon 
and the $^{19}$F nucleus. 
 The Cu$^{2+}$ electronic moments, which dominate the spectra for 
$T<T_{\mathrm{N}}$,
 are no longer ordered in the paramagnetic regime,
and fluctuate very rapidly
on the muon time scale. They are therefore motionally narrowed
from the spectra, leaving the muon sensitive to the quasistatic nuclear 
magnetic moments. The signal arising from F--$\mu^{+}$ states persist in these
materials to temperatures well above 100~K. 

In fluorinated materials, where the muon-spin is relaxed through
interaction 
with
nuclear moments, we expect two contributions to the $\mu^{+}$SR
spectra. 
The first is from muons that
strongly couple to fluorine nuclei, which give rise to contributions
derived from Eq.~(\ref{polarization}). Note that since our
measurements
are carried out on powder samples, $D_{z}(t)$ also includes the
effect of angular averaging.
The second contribution is from those muons
weakly coupled to a large number of nuclei, 
leading to Gaussian relaxation in the LMF 
model, as described above.
Above the critical temperature $T_{\mathrm{N}}$,
spectra were found to be well described by the resulting
polarization function
\begin{equation}
A(t) = A_{0} [p_{1} D_{z}(t) \exp(-\lambda t) + p_{2}
\exp(-\sigma^{2}t^{2})] +A_{\mathrm{bg}}\label{fit},
\end{equation}
where  $A_{0}$ is the signal arising from the
sample, $p_{1}+p_{2}=1$, and $A_{\mathrm{bg}}$ accounts for those muons that stop in
the sample holder or cryostat tails. 

Below we discuss experimental results\cite{experiment} for 
three molecular materials where stopping states arise
that are different from the conventional F--$\mu$--F state
and exemplify three distinct classes of implanted muon state.

{\it Case I: Interaction with a single fluorine.}
The compound CuF$_{2}$(H$_{2}$O)$_{2}$(pyz) \cite{manson} is formed 
from CuF$_{2}$O$_{2}$N$_{2}$ octahedra linked
with pyrazine bridges along the $a$-direction 
to form linear chains (Fig.~\ref{cuf2}(a)). Extensive
hydrogen bonding interactions tether the chains into the
3D network shown in Fig.~\ref{cuf2}(b).
The material undergoes an AFM transition at
$T_{\mathrm{N}}=2.54(8)$~K below which oscillations in the
muon asymmetry are observed \cite{manson}.  
ZF $\mu^{+}$SR spectra measured on CuF$_{2}$(H$_{2}$O)$_{2}$(pyz) 
above $T_{\mathrm{N}}$ are shown in Fig.~\ref{cuf2}(b)
where we see slow, heavily damped oscillations characteristic
of F--$\mu^{+}$ dipole coupled states.
The spectra are most successfully modelled by assuming the muon is strongly
coupled with a single $I=\frac{1}{2}$ F spin, localized a
distance $d=0.110(2)$~nm away from a fluorine nucleus. 
The resulting energy level structure and allowed transitions for this
scenario are shown inset in Fig.~\ref{cuf2}(c).
The F--$\mu^{+}$ spin system consists of three distinct
energy levels with three allowed transitions between them 
giving rise to the distinctive three-frequency oscillations observed
(inset Fig.~\ref{cuf2}c). 
The signal is described by a
polarization function 
$
D_{z}(t) = \frac{1}{6} 
\left[ 1 + \sum_{j=1}^{3} u_{j} \cos (\omega_{j} t) \right],
$
where 
$u_{1}=2$, $u_{2}=1$ and $u_{3}=2$.
The transition frequencies (shown in Fig.~\ref{cuf2}(c)) are given by 
$\omega_{j}= j \omega_{\mathrm{d}}/2$
where $\omega_{\mathrm{d}}=\mu_{0} \gamma_{\mu} \gamma_{\mathrm{F}}/4 \pi r^{3}$,
and $r$ is the F--$\mu^{+}$ separation. 
The resulting fit is shown in Fig.~\ref{cuf2}(c) using the parameters
listed in Table \ref{table}. 

\begin{table}
   \caption{Parameters for Eq.(\ref{fit}) for each material studied.
with resulting $\chi^{2}$ for the fits
described in the main text. $\chi^{2}_{\mathrm{F \mu F}}$
corresponds to the best fits obtained using a 
conventional F--$\mu^{+}$--F model for comparison.
\label{table}}
     \begin{ruledtabular}
   \begin{tabular}{lllllll}
                               & $p_{1}$  & $\lambda$ & $p_{2}$ &
                               $\sigma$ &$\chi^{2}$ &  $\chi^{2}_{\mathrm{F
                                 \mu F}}$\\\hline
CuF$_{2}$(H$_{2}$O)$_{2}$(pyz)     & 0.43    & 0.36   &  0.57  &   0.56 & 1.2 &1.6\\
CuNO$_{3}$(pyz)$_{2}$PF$_{6}$      & 0.65     & 0.24   & 0.35   &   0.45 & 2.2&4.1\\
$[$Cu(HF$_{2}$)(pyz)$_{2}]$ClO$_{4}$ & 0.73     & 0.24  &  0.27  &
0.43 & 2.2  & 3.9\\
   \end{tabular}
     \end{ruledtabular}
\end{table}

We note first that the strong interaction of the muon and a single F
is unusual. However,
a muon stopping site between
two fluorines is probably made energetically unfavourable due to the
presence of the protons on the H$_{2}$O groups 
which are hydrogen bonded to the fluorines (and stabilize the solid).
In this material the smallest F--F distance (between adjacent chains) 
is 0.34~nm and a position midway between these 
two fluorines invariably lies very close ($\sim 0.13$~nm) to the
protons on a H$_{2}$O group.
We further exclude a muon
site in CuF$_{2}$(H$_{2}$O)$_{2}$pyz
where the muon is bonded to F but sits near the protons on
the H$_{2}$O groups since fits to such a configuration are not able to
account for the measured data. It is more probable that the muon's
separation from the F ion is not precisely
in the $a$-$b$ plane, but rather has a small
component in the $c$ direction, taking it closer to the electron
density on the aromatic rings. 
We note further that we have also observed coupling of the muon to a single
F in the polymer PVDF (--(CH$_{2}$CF$_{2}$)$_{n}$--) below the
polymer glass transition \cite{me_unpublished}. This demonstrates that this stopping state
is not unique.

{\it Case II: Crooked F-F bond with the PF$_{6}^{-}$ ion.}
The quasi 2D compound [Cu(NO$_{3}$)(pyz)$_{2}$]PF$_{6}$ 
\cite{turnbull,woodward} is formed
from infinite 2D sheets of [Cu(pyz)$_{2}$]$^{2+}$ lying in the $ab$
plane. These are
linked along the $c$-direction by NO$_{3}^{2-}$ ions. The
PF$_{6}^{-}$ anions occupy the body-centred positions within the
pores. The PF$_{6}^{-}$ anion has a regular octahedral geometry with a P--F
distance of 0.157~nm.

\begin{figure}
\begin{center}
\epsfig{file=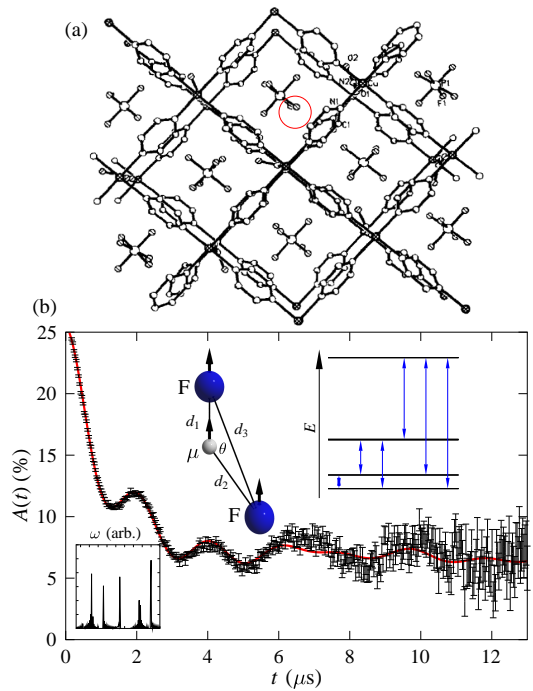,width=7.cm}
\caption{(Color online) (a) Structure of [Cu(NO$_{3}$)(pyz)$_{2}$]PF$_{6}$
in the $a$-$b$ plane showing 2D sheets of [Cu(pyz)$_{2}$]$^{2+}$
and PF$_{6}^{-}$ anions occupying the pores in the structure \cite{turnbull}.
 The circled region shows 
sample fluorines near which the muon is expected to localize.
(b) Example ZF spectra measured at $T=5.7$~K with
a fit reflecting the configuration shown. {\it Inset:} proposed
F--$\mu^{+}$ configuration,
 (top right)
the resulting energy levels with
allowed transitions and (bottom left) the calculated  frequency spectrum.
\label{pf6}}
\end{center}
\end{figure}

Our measurements show that this
material magnetically orders below $T_{\mathrm{N}}=2.0(2)$~K. 
Above $T_{\mathrm{N}}$
we again observe a signal from dipole coupling. 
An example spectrum measured above $T_{\mathrm{N}}$ is shown in 
Fig.~\ref{pf6}. Although one would expect the muon to localise near a PF$_{6}$ 
anion, it is not clear {\it a priori} how many fluorine
centres will be strongly coupled to the muon spin. Modelling the
spectra reveals that two fluorine spins interact with the $\mu^{+}$, 
but not via the linear F--$\mu^{+}$--F. 
Instead we find the F--$\mu^{+}$--F bond angle to be $\theta=143(1)^{\circ}$
with F--$\mu^{+}$ lengths $d_{1}=0.106(3)$~nm and $d_{2}=0.156(3)$~nm. 
The
F--F distance, which is 0.22~nm in the unperturbed material, lengthens
to 0.25~nm. 
This configuration is shown {\it inset} in Fig.~\ref{pf6}. A
fit to Eq.~\ref{fit} resulted in the parameters listed in
Table~\ref{table}
and is shown in Fig.~\ref{pf6}. 
It should also be expected that other materials containing $X$F$_{6}^{-}$ ions
($X=$Sb or As) will have stopping states of this type.

{\it Case III: Interaction with the HF$_{2}^{-}$ ion.}
The series of coordination polymers [Cu(HF$_{2}$)(pyz)$_{2}$]$Y$ 
are formed from infinite 2D sheets of
[Cu(pyz)$_{2}$]$^{2+}$ in the $ab$ plane (as in the case of 
[Cu(NO$_{3}$)(pyz)$_{2}$]PF$_{6}$ above). These are connected along
the $c$-axis by linear HF$_{2}^{-}$ anions to form a pseudo-cubic
network. Small tetrahedral or octahedral anions $Y$
occupy the body centred positions in the
pseudo-cubic pores \cite{manson2}. 

\begin{figure}
\begin{center}
\epsfig{file=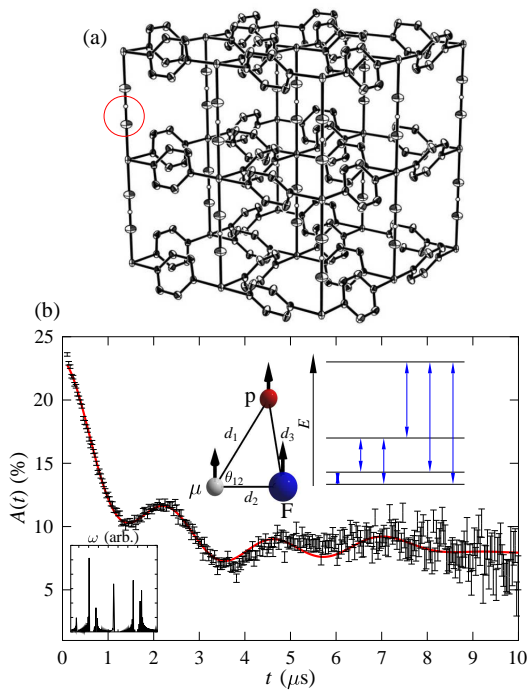,width=7.cm}
\caption{(Color online)
(a) Crystal structure if [Cu(HF$_{2}$)(pyz)$_{2}$]$X$, with
$X$ atoms omitted for clarity. The circled region shows a
sample fluorine and proton near which the muon is expected to localize.
(b) Example ZF spectra measured at $T=11$~K with
a fit reflecting the configuration shown. {\it Inset:} proposed
F--$\mu^{+}$--p  configuration, (top right)
the resulting energy levels with
allowed transitions and (bottom left) the calculated frequency spectrum.
\label{hf2}}
\end{center}
\end{figure}

An example spectrum measured above $T_{\mathrm{N}}=1.94$~K for
[Cu(HF$_{2}$)(pyz)$_{2}$]ClO$_{4}$
is shown in 
Fig.~\ref{hf2}(b). In this case we would expect the muon to localize
near the HF$_{2}^{-}$ anion. The spectrum obtained is qualitatively similar to
that expected for the interaction of a muon with a single fluorine
ion. However, the best fit is obtained if a third spin centre is
included. For the tightly bound HF$_{2}$ ion we might expect this
third centre to be that of the nearest proton (see {\it inset}
Fig.~\ref{hf2}(b)). This configuration is
found to fit the data successfully yielding a F--$\mu^{+}$ distance of
$d_{1}=0.111(3)$~nm, a $\mu^{+}$--$p$ distance of $d_{2}=0.161(3)$~nm and a 
F--$\mu^{+}$--p angle of $\theta= 57(1)^{\circ}$. The p--F
distance $d_{3}$, which is 0.11~nm in the unperturbed system, is found to
be increased to $d_{3}=0.137(3)$~nm in the presence of the muon. The
resulting fit to Eq.~(\ref{fit}) 
results in the fitting parameters given in Table \ref{table}
and is shown in Fig.~\ref{hf2}(b).

In addition we note that no dipole-dipole signal is detected
from muons stopping near the BF$_{4}^{-}$ ion; measurements on
Cu(pyz)$_{2}$(BF$_{4})_{2}$ and Cu(C$_{5}$H$_{6}$NO)$_{6}$(BF$_{4}$)$_{2}$
\cite{me_unpublished} do not show a F--$\mu^{+}$ dipole-dipole signal, 
while measurements on  [Cu(HF$_{2}$)(pyz)$_{2}$]BF$_{4}$
\cite{manson2} show the signal from the HF$_{2}^{-}$ ion described
above. 

In the conventional F--$\mu^{+}$--F case the F atoms
may each shift by large distances ($\sim 1$~\AA) from their equilibrium
positions towards the $\mu^{+}$ \cite{brewer},  demonstrating that the muon
introduces a non-negligible local distortion in the material. 
If however, by analogy with the
$V_{\mathrm{k}}$ defect center in alkali halides \cite{hayes}, 
 the F--$\mu^{+}$--F complex acts like an independent molecule in the 
crystal,
the distortion
in the other ion positions will be much less significant
than the distortion of the two F$^{-}$ ions. The $V_{\mathrm{k}}$ centre
analogy will be less accurate in cases I-III described above, where
the non-linear bonds demonstrate that the muon stopping state cannot
be regarded as separate from its surroundings. In these cases the
tightly bound nature of the fluorine-containing complexes prevents 
an independent quasi-molecular impurity from forming. 

We have presented three examples of the F--$\mu^{+}$ stopping states
in molecular materials. The inherently quantum mechanical
interaction of the muon spin leads to entangled states
involving the surrounding nuclei and has
allowed a characterization of the muon stopping state and facilitates
a quantitative estimate of the degree to which the muon distorts 
its surroundings. These results demonstrate that the introduction of
fluorine ions in molecular magnets can provide ``traps'' for muons,
so that the local spin structure in such systems can be probed from
well characterized muon sites.

Part of this work was carried out at the ISIS facility, Rutherford Appleton
Laboratory, UK. This work is supported by the EPSRC (UK).
T.L. acknowledges support from the Royal Commission for the Exhibition
of 1851. J.L.M. acknowledges an award from Research Corporation.
Work at
Argonne National Laboratory is sponsored by the U. S. Department
of Energy, Office of Basic Energy Sciences, Division of Materials
Sciences, under Contract DE-AC02-06CH11357.
We thank A.M. Stoneham for useful discussions.

\end{document}